\documentclass[12pt,a4paper]{scrartcl}
\pdfoutput=1
\usepackage{amsmath,amssymb}
\usepackage{subfigure}

\usepackage{hyperref}
\usepackage{graphicx}
\usepackage{feynmf} 
\addtokomafont{disposition}{\rmfamily\boldmath}
\let\tmptitle\title\renewcommand{\title}[1]{\tmptitle{\LARGE #1}}
\let\tmpauthor\author\renewcommand{\author}[1]{\tmpauthor{\large #1}}
\let\tmpdate\date\renewcommand{\date}[1]{\tmpdate{\normalsize #1}}
\newcommand{\abstrct}[1]{\begin{abstract}\vspace{-2em}\small\noindent#1\end{abstract}}

\title{\Large
Freed Leptogenesis
}
\date{\today}                                                                                                                                                                                                                                                                                                                                                                                  
\author{
Dmitry~V.~Zhuridov
\footnote{Email: dmitry.zhuridov@sns.it}
\\ \normalsize\itshape
Scuola Normale Superiore, and INFN, Piazza dei Cavalieri 7, 56126 Pisa, Italy 
}

\begin{document}

\maketitle

\abstrct{%
Economical extensions of the Standard Model (SM), in which famous Davidson--Ibarra bound on the $CP$ asymmetry 
relevant for leptogenesis may be significantly relaxed by the loop effects, comparing to predictions of the SM extended only by 
heavy right-handed neutrinos with hierarchical masses, are discussed. This leads to decreasing of the lower bound on the heavy neutrino masses 
and increasing of the upper bound on the light neutrino masses, which is testable. In addition, the considered theory may help to solve the dark matter problem.
}

\begin{fmffile}{bkz}

\section{Introduction}

The observable small nonzero neutrino masses and baryon asymmetry of the Universe (BAU)~\cite{PDG2010} provide strong evidences of physics beyond the Standard Model (SM).
The see-saw mechanism~\cite{seesawA,seesawB,seesawC,seesawD,seesawE,seesawF} gives economical explanation of the lightness of neutrinos by adding the heavy Majorana neutrinos to the SM particle content,
which generate the small neutrino masses by the tree level perturbative interaction with the SM Higgs vacuum expectation value (VEV).
In addition, the BAU may be explained by generating the lepton asymmetry in the out-of-equilibrium decays of these heavy neutrinos 
and converting it to a baryon asymmetry by sphaleron transitions~\cite{KRS} in the usual baryogenesis~\cite{Sakharov} via leptogenesis (LG) scenario~\cite{Fukugita_Yanagida}.
However the successful LG in this simple SM extension requires (in the case of hierarchical heavy neutrinos) a strong upper bound on the relevant $CP$ asymmetry, 
which was introduced in~\cite{DI0A,DI0B,DI0C,DI}, and is called Davidson--Ibarra (DI) bound. 
This results also in the lower bound on the right-handed neutrino masses of $\sim10^9$~GeV~\cite{DI} 
and the upper bound on the left-handed neutrino masses of $\sim0.1$~eV~\cite{0302092,Strumia}. 
By generalizing the SM to the Minimal Supersymmetric Model the bound on the $CP$ asymmetry is increasing only by the factor of two, 
which leads to the famous gravitino problem~\cite{Khlopov1,Balestra,Khlopov2}.
However in the case of quasi-degenerate heavy neutrinos a resonant enhancement of the $CP$ asymmetry may happen~\cite{PilaftsisA,PilaftsisB}.

Another possible solution for the problem of small observable neutrino masses is its radiative 
generation~\cite{ZeeA,ZeeB,ZeeC,ZeeD,Ma_model,Perez-Wise_modelA,Perez-Wise_modelB,Perez-Wise_modelC}.
In this paper we consider generation of the neutrino masses at both tree and loop levels. We show that the theories with 
analytical relation between the couplings relevant for tree and loop contributions to the neutrino masses may significantly relax the DI bound 
in the case when these tree and loop terms approximately cancel each other.
As a result, strongly hierarchical heavy neutrino masses in this theory may be tested at the Large Hadron Collider (LHC) and next particle 
facilities~\cite{heavy_neutrinosA,heavy_neutrinosB,ABZ11,ABZ12,ABZ21,ABZ22,ABZ3}.
The discussed analytical relation may come from the structure of grand unified theories (GUT)~\cite{Georgi_Glashow}, 
in which the particles involved in the tree and loop contributions to the neutrino masses belong to the same multiplets.
In particular, Renormalizable Adjoint $SU(5)$ model~\cite{0702287} is one of the minimal realistic GUTs, in which a linear relation of this type is realized~\cite{KZ}.

In the next section we investigate generation of the neutrino masses in the economical SM extensions with the loop contribution to the neutrino masses 
analogous to~\cite{Ma_model} and \cite{Perez-Wise_modelA,Perez-Wise_modelB,Perez-Wise_modelC}.
We analyze LG in section~\ref{section:LG}, and conclude in section~\ref{section:summary}.

\section{Generation of neutrino masses}\label{section:nu_masses_generation}

\begin{figure}[h]
\centering
\subfigure[]{
\fmfframe(1,6)(0,-10){
\begin{fmfgraph*}(30,25)
\fmfleft{i1}
\fmfright{o1}
\fmftop{t1,t2,t3,t4}
\fmf{fermion,label=$\nu_{L\alpha}$,label.side=right}{i1,v1}
\fmf{fermion,tension=2}{v1,v3}
\fmf{fermion,tension=2}{v2,v3}
\fmfv{label=$N_i$,label.angle=-90}{v3}
\fmf{fermion,label=$\nu_{L\beta}$,label.side=left}{o1,v2}
\fmffreeze
\fmf{scalar}{t2,v1}
\fmf{scalar}{t3,v2}
\fmflabel{$\langle \phi\rangle$}{t2}
\fmflabel{$\langle \phi\rangle$}{t3}
\end{fmfgraph*}
} }
\subfigure[]{
\fmfframe(1,6)(0,-10){
\begin{fmfgraph*}(30,25)
\fmfleft{i1}
\fmfright{o1}
\fmftop{t1,t2,t3,t4}
\fmf{fermion,label=$\nu_{L\alpha}$,label.side=right}{i1,v1}
\fmf{fermion,label=$\nu_{L\beta}$,label.side=left}{o1,v1}
\fmffreeze
\fmf{scalar}{t2,v1}
\fmf{scalar}{t3,v1}
\fmflabel{$\langle \phi\rangle$}{t2}
\fmflabel{$\langle \phi\rangle$}{t3}
\fmfblob{.2w}{v1}
\end{fmfgraph*}
} }
\caption{Considered contributions to the neutrino masses. The arrows in fermionic lines show flow of lepton number. 
}
\label{fig:1:nu:masses}
\end{figure}
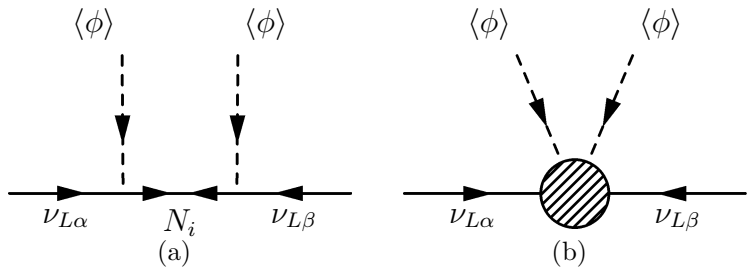

Consider theory with  the neutrino masses generated by heavy Majorana fermions $N_i$,\footnote{The correspondent 
mechanism of generation of the neutrino masses is called type I or type III see-saw, 
depending on whether $N_i$ is singlet or triplet fermion, respectively.} as shown in Fig.~\ref{fig:1:nu:masses} $a$, 
and by other new heavy particles, which is shown effectively in Fig.~\ref{fig:1:nu:masses} $b$ after integration out these particles.
It is well known that besides generating the neutrino masses the heavy fermions $N_i$ 
can be at the same time responsible for the LG. 
In the case when the new heavy particles, involved in the contribution in Fig.~\ref{fig:1:nu:masses} $b$, are decoupled from LG this contribution may relax the connection 
between the neutrino masses and LG, namely the DI bound.
Such {\it LG} we call {\it Freed}. 

\begin{figure}[h]
\centering
\fmfframe(1,6)(0,-10){
\begin{fmfgraph*}(40,35)
\fmfleft{i1}
\fmfright{o1}
\fmftop{t1,t2,t3,t4}
\fmf{fermion,label=$\nu_{L\alpha}$,label.side=right}{i1,v1}
\fmf{fermion}{v1,v2}
\fmf{fermion}{v3,v2}
\fmf{fermion,label=$\nu_{L\beta}$,label.side=left}{o1,v3}
\fmfv{label=$N$,label.angle=-90}{v2}
\fmffreeze
\fmf{phantom,left=0.5,tension=0.5}{v1,v4,v3}
\fmf{scalar,label=$\eta$,label.side=right,right=0.5,tension=0.5}{v4,v1}
\fmf{scalar,label=$\eta$,label.side=left,left=0.5,tension=0.5}{v4,v3}
\fmf{scalar}{t2,v4}
\fmf{scalar}{t3,v4}
\fmflabel{$\langle \phi\rangle$}{t2}
\fmflabel{$\langle \phi\rangle$}{t3}
\end{fmfgraph*}
}
\caption{Possible 1-loop contribution to neutrino masses. The arrows in fermionic lines show flow of lepton number.}
\label{fig:2:loop:nu:masses}
\end{figure}
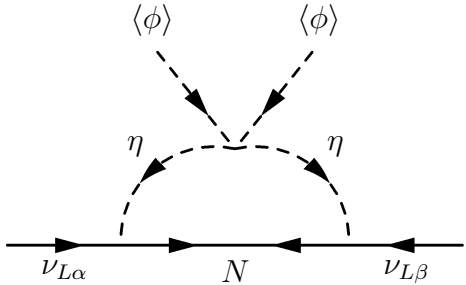

In this paper we discuss a particular class of theories with the dominant 1-loop contribution to the effective vertex in Fig.~\ref{fig:1:nu:masses} $b$, 
shown in Fig.~\ref{fig:2:loop:nu:masses}, where $N$ is new $SU(2)_L$ singlet Majorana fermion and $\eta$ is new $SU(2)_L$ doublet scalar. 
The two possible classes of models, which generate this contribution, 
were introduced by Ma~\cite{Ma_model} and Perez--Wise~\cite{Perez-Wise_modelA,Perez-Wise_modelB,Perez-Wise_modelC}. Consider extensions of this models by several singlet or singlet and triplet Majorana fermions $N_i$. 
In the minimal case we need only two $N_i$ for non-degenerate neutrino masses and successful LG.
The new particles in these extended Ma (EMM) and Perez--Wise (EPWM) models with their properties under the SM groups $G_\text{SM}=SU(3)_c\times SU(2)_L\times U(1)_Y$ 
and the discrete symmetry $Z_2$ (in the case of EMM) are listed in Tables~\ref{table:Ma} and \ref{table:Perez-Wise}, respectively. All the SM particles in EMM have positive $Z_2$ parity. 
We take this definitions of extended models for two reasons:  simplicity in the case of EMM, and reproduction of particles 
responsible for the neutrino masses and LG of Adjoint $SU(5)$~\cite{0702287,KZ,LG_BP1,LG_BP2} in the case of EPWM.
\footnote{Important for LG is weather the lightest Majorana fermion $N_1$ is $SU(2)$ singlet or triplet. According to this, in general,
both {\it singlet} and {\it triplet} types of extensions can be considered for Ma model, and same for Perez--Wise model. 
However 
in our definitions EMM (EPWM) generates singlet (triplet) LG.}

\begin{table}[htdp]
\caption{Quantum numbers of the new particles in EMM under $G_{\text{SM}}\times Z_2$.}
\label{table:Ma}
\begin{center}
\begin{tabular}{|c|c|c|c|c|}
	\hline
    Field & $N_i$ & $N$ & $\eta$ \\ 
    \hline
     $Z_2$ & $+$ & $-$ & $-$ \\ 
    $SU(3)_c$ & 1 & 1 & 1 \\
    $SU(2)_L$ & 1 & 1 & 2 \\ 
      $U(1)_Y$ & 0 & 0 & $1/2$ \\  
    \hline
\end{tabular}
\end{center}
\label{default}
\end{table}

\begin{table}[htdp]
\caption{Quantum numbers of the new particles in EPWM and corresponding particles in Adjoint $SU(5)$ under $G_{\text{SM}}$.}
\label{table:Perez-Wise}
\begin{center}
\begin{tabular}{|c|c|c|c|c|}
	\hline
    Field & $N_1\equiv\rho_3$ & $N_2\equiv\rho_0$ & $N\equiv\rho_8$ & $\eta\equiv S_8$ \\ 
    \hline
     $SU(3)_c$ & 1 & 1 & 8 & 8 \\ 
    $SU(2)_L$ & 3 & 1 & 1 & 2 \\ 
      $U(1)_Y$ & 0 & 0 & 0 & $1/2$ \\  
    \hline
\end{tabular}
\end{center}
\label{default}
\end{table}

\subsection{Generation of tree and loop terms}

The most general renormalizable CP conserving scalar potential in EMM is analogous to the one in the inert doublet model~\cite{Deshpande-Ma,Barbieri-Hall-Rychkov,1106.1719}
\begin{eqnarray}
 V =  \mu_1^2 |\phi|^2 + \mu_2^2 |\eta|^2 + \lambda_1 |\phi|^4 + \lambda_2 |\eta|^4 + \lambda_3 |\phi|^2|\eta|^2 
+ \lambda_4 |\phi^\dag\eta|^2 
+ \frac{\lambda_5}{2} \left[ (\phi^\dag\eta)^2 + \text{H.c.} \right],
\end{eqnarray}
where $\mu_i^2$ and $\lambda_i$ are real, $\phi$ is the SM Higgs doublet. (The most general renormalizable scalar potential in EPWM and the bounds on its couplings are discussed in Refs.~\cite{Manohar-Wise,He:2013tla}, where the coupling $\lambda_3$ plays the role of $\lambda_5$ in Eq.~(1), assuming the tracing of color matrices.) 
The Higgs boson mass squared is $M_h^2=4\lambda_1v_0^2$, where $v_0=174$~GeV is the Higgs VEV. The inert doublet $\eta$ has zero VEV. 
With positive squared masses of scalars this potential is 
bounded from below if and only if~\cite{Ivanov}
\begin{eqnarray}
 \lambda_{1,2}>0, \quad \lambda_3 > -\sqrt{\lambda_1\lambda_2}, \quad |\lambda_5| < 2\sqrt{\lambda_1\lambda_2} + \lambda_3 + \lambda_4.
\end{eqnarray}

The Yukawa and mass terms in the Lagrangian relevant for the neutrino masses can be written as
\begin{eqnarray}\label{eq:Lagrangian}
 -\cal{L}_\nu &=& Y_{i\alpha}\bar L_\alpha N_i\phi + \frac{1}{2}\bar N_iM_{ij}N_j^c \nonumber\\
  &+&  h_\alpha \bar L_\alpha N\eta + \frac{1}{2}\bar NM_NN^c + \text{H.c.},
\end{eqnarray}
where $\alpha$ is flavor index, $L=(e_L,\nu_L)^T$ is the SM lepton doublet, and proper contractions of the $SU(2)_L$ and color indexes should be done in EPWM. 
By integrating out $N_i$ and calculating the loop in Fig.~\ref{fig:2:loop:nu:masses} we have
\begin{eqnarray}
 \cal{L}_\nu^\text{eff} &=& \frac{1}{2} \bar L\phi \left( Y^T M^{-1} Y \right) \phi^T L^c 
+ \frac{1}{2\Lambda} \bar L\phi \left(hh^T\right) \phi^T L^c + \text{H.c.}
\end{eqnarray}
In the mass basis of $N_i$, $M=\text{diag}(M_1,M_2)\equiv D_M$, after absorption of the minus sign by rotation of $\nu_{\alpha L}$ fields 
the neutrino mass matrix can be written as
\begin{eqnarray}
 M_\nu = v_0^2 Y^TD_M^{-1}Y + \frac{v_0^2}{\Lambda} hh^T 
 \equiv  M_\nu^\text{tree} + M_\nu^\text{loop},  \label{eq:nu_masses:1}
\end{eqnarray}
where $M_\nu^\text{tree}$ is type I~\cite{seesawA,seesawB,seesawC,seesawD,seesawE,seesawF} (type I plus III~\cite{seesawIIIA,seesawIIIB}) see-saw contribution in EMM (EPWM), 
and $\Lambda$ is the high-energy mass scale, generated in loop, which may be positive or negative, depending on the relevant couplings.
For the loop, shown in Fig.~\ref{fig:2:loop:nu:masses},
\begin{eqnarray}\label{eq:Lambda}
 \Lambda = \frac{16\pi^2}{C\lambda_5} F^{-1}\left( \frac{M_\eta}{M_N} \right) M_N 
\simeq \frac{8\pi^2}{C\lambda_5} \left( \ln\frac{M_N}{M_\eta} - \frac{1}{2} \right)^{-1} M_N \quad \text{for} \ \ M_\eta\ll M_N,
\end{eqnarray}
where $C=1$ in EMM and $C=N_c^2-1=8$ ($N_c=3$ is the number of colors) in EPWM, and the loop function
\begin{eqnarray}
 F(x)&=& \frac{x^2-1-\ln x^2}{(1-x^2)^2} \\
 &=&-(1+2\ln x) + {\mathcal O}(x^2\ln x) \quad\quad {\rm for} \ \ x\ll1  \nonumber
\end{eqnarray}
comes from the finite part of the Passarino--Veltman function $B_0$~\cite{Passarino-Veltman,Zapata}.
We note that the only difference between the loop contributions to the neutrino masses in EMM and EPWM is encoded 
by the factor $C$ in Eq.~\eqref{eq:Lambda}. 

The difference between the new physics contributions of neutral and charged current processes at low energies is measured by the $T$ parameter~\cite{PDG2010}, 
which is constrained by the present experiments as $T=0.05\pm0.11$~\cite{T_parameter}.
The contribution of the fermionic triplet to $T$ is zero in the case of 
mass degeneracy of its neutral and charged components, because the contributions to self-energies of the Goldstones $\phi^+$ and $\chi$ 
(see Appendix C in \cite{Barbieri-Hall-Rychkov} for $\Delta\rho$, which is proportional to $T$) cancel each other.

\subsection{Connection of tree and loop terms}

Consider EMM or EPWM as a part of more general theory, which possesses analytical relation among the Yukawa couplings $Y$ and $h$ in Eq.~\eqref{eq:Lagrangian}. 
For simplicity, let it be a linear relation
\begin{eqnarray}\label{eq:linear_relation}
 h_\alpha^T = a_iY_{i\alpha}
\end{eqnarray}
with real $a_i$. In particular, in Adjoint $SU(5)$ model
\begin{eqnarray}\label{eq:a_i}
 a_1 = \mp \frac{1}{16\sqrt{6}} \frac{v_0}{|v_{45}|}, \quad a_2 = \pm \frac{\sqrt{5}}{24\sqrt{2}} \frac{v_0}{|v_{45}|},
\end{eqnarray}
where $v_{45}$ is the VEV of ${\bf45}_H$ representation.
Because the same contraction of the representations $\bar{\bf5}$-${\bf24}$-${\bf45}_H$ contains the terms, 
which contribute to both tree and loop level neutrino masses~\cite{KZ}. More explicitly, 
the term $p_\alpha \bar {\bf5}_\alpha\, {\bf24}\, {\bf45}_H$ in the Lagrangian, which is involved in the generation of 
the loop neutrino mass term proportional to $p_\alpha p_\beta$, 
generates also type I and III see-saw contributions to the neutrino masses, which are dependent on the same coefficient $p_\alpha$.
Notice that in Eq.~\eqref{eq:a_i} $|a_i|\lesssim1$.

Using Eq.~\eqref{eq:linear_relation}, Eq.~\eqref{eq:nu_masses:1} could be rewritten as
\begin{eqnarray}
 M_\nu = v_0^2 Y^T{\cal{M}}^{-1}Y, 
  \label{eq:nu_masses:2}
\end{eqnarray}
where in the case of two $N_i$
\begin{eqnarray}\label{eq:matrix_M}
 \mathcal{M}^{-1} = \left(
\begin{array}{cc}
M_{1}^{-1}+a_1^2\Lambda^{-1} & a_1a_2\Lambda^{-1} \\
a_1a_2\Lambda^{-1} & M_{2}^{-1}+a_2^2\Lambda^{-1} \\
\end{array}
\right).
\end{eqnarray}

The neutrino mass matrix in Eq.~(\ref{eq:nu_masses:2}) can be rewritten as
\begin{eqnarray}\label{eq:nu_masses:diag}
 M_\nu = v_0^2\, Y^{\prime T} D_\mathcal{M}^{-1}Y^{\prime},
\end{eqnarray}
by using the orthogonal transformation
\begin{eqnarray}
 Q \, Y &=& Y^\prime, \label{eq:Yprime} \\
  Q\,\mathcal{M}\,Q^T &=& D_\mathcal{M}, 
\end{eqnarray}
where
\begin{eqnarray}\label{eq:Mmatrix}
 \mathcal{M} = \frac{M_1 M_2 \Lambda}{\Lambda + a_1^2M_1 + a_2^2M_2} \left(
\begin{array}{cc}
M_{2}^{-1}+a_2^2\Lambda^{-1} & -a_1a_2\Lambda^{-1} \\
-a_1a_2\Lambda^{-1} & M_{1}^{-1}+a_1^2\Lambda^{-1} \\
\end{array}
\right)
  \equiv  M_0^2\left(
\begin{array}{cc}
a & c \\
c & b \\
\end{array}
\right)
\end{eqnarray}
is modified mass matrix of heavy fermions,
$D_\mathcal{M} = \text{diag}(\tilde M_{1},\tilde M_{2})$ with the eigenvalues 
\begin{eqnarray}\label{eq:tildM12}
 \tilde M_{1,2} = \frac{M_0^2}{2}\left( a+b \mp \sqrt{(b-a)^2+4c^2} \right)
= \frac{M_0^2}{2}\left( a+b \mp \sqrt{(a+b)^2-4M_0^{-2}} \right)
\end{eqnarray}
and
\begin{eqnarray}
 Q = \left(
\begin{array}{cc}
\cos q & \sin q \\
-\sin q & \cos q \\
\end{array}
\right)
\end{eqnarray}
is real orthogonal matrix with the mixing
\begin{eqnarray}\label{eq:sinq}
 \sin q = - \frac{\sqrt{2}c}{\sqrt{(b-a)[(b-a)+\sqrt{(b-a)^2+4c^2}]+4c^2}}.
\end{eqnarray}
Eqs.~\eqref{eq:Mmatrix} and \eqref{eq:tildM12} show that $\tilde M_2$ has singularity at $\Lambda= -a_1^2M_1 -a_2^2M_2$.


For hierarchical $N_i$ with $M_1 \ll \min(|\Lambda|, M_2)$ we have following approximations 
\begin{eqnarray}\label{eq:Mi_approx}
 \tilde M_1 \simeq M_1 \left( 1-a_1^2\frac{M_1}{\Lambda} \right),  \qquad   \tilde M_2 \simeq \frac{M_2\Lambda}{\Lambda + a_2^2M_2}
\end{eqnarray}
and
\begin{eqnarray}\label{eq:sinq_approx}
 \sin q \simeq a_1a_2 \frac{M_1}{\Lambda},  \qquad  \cos q \simeq 1-\frac{a_1a_2M_1}{2\Lambda}.
\end{eqnarray}

\subsection{Parametrization of neutrino masses}

For explanation of the neutrino experimental data we use the standard Casas-Ibarra~\cite{Casas-Ibarra} parametrization of the Yukawa couplings $Y^\prime$ as
\begin{eqnarray}\label{eq:Casas-Ibarra}
 Y^\prime=v_0^{-1}D_\mathcal{M}^{1/2}\Omega D_\nu^{1/2}U^\dag,
\end{eqnarray}
where $\Omega$ is complex orthogonal (or partly orthogonal for the number of $N_i$ different from three) matrix, and $U$ is the PMNS lepton mixing matrix, which diagonalizes the neutrino mass matrix 
in the flavor basis according to
\begin{eqnarray}\label{eq:PMNS}
  U^T M_\nu U = {\rm diag}(m_1,m_2,m_3) \equiv D_\nu.
\end{eqnarray}

In the case of two $N_i$ one of the light neutrinos is massless. Hence the quasi-degenerate neutrinos are forbidden, and the only allowed neutrino mass spectra are
\begin{itemize}
\item Normal Hierarchical (NH)
\begin{eqnarray}
 m_1 = 0, \quad m_2 = \sqrt{\Delta m_{\text{sol}}^2}, \quad m_3 = \sqrt{\Delta m_{\text{atm}}^2};
 \label{eq:NH}
\end{eqnarray}
\item Inverted Hierarchical (IH)
\begin{eqnarray}
 m_3=0, \quad m_1 = \sqrt{\Delta m_{\text{atm}}^2-\Delta m_{\text{sol}}^2}, \quad m_2 = \sqrt{\Delta m_{\text{atm}}^2};
\end{eqnarray}
\end{itemize}
where $\Delta m_{\text{sol}}^2 = 7.65\times10^{-5}$~eV$^2$ and $\Delta m_{\text{atm}}^2 = 2.40\times10^{-3}$~eV$^2$ 
are the mass-squared differences of solar and atmospheric neutrino oscillations~\cite{PDG2010}.
In this case $\Omega$ is $2\times3$ matrix, which can be written as~\cite{Ibarra-Ross}
\begin{eqnarray}
 \Omega^{\text{NH}} = 
\left(
 \begin{array}{ccc}
  0 & \cos z & \pm\sin z \\
  0 & -\sin z & \pm\cos z \\
 \end{array}
\right), \qquad
 \Omega^{\text{IH}} = 
\left(
 \begin{array}{ccc}
  \cos z & \pm\sin z & 0 \\
  -\sin z & \pm\cos z & 0 \\
 \end{array}
\right)
\end{eqnarray}
in the normal and inverted hierarchy, respectively; where $z$ is the complex angle.


\section{Leptogenesis}\label{section:LG}

\subsection{$CP$ asymmetry}
\subsubsection{General formulas}

The $CP$ asymmetry is generated in the decays of $N_i$. Relevant for the unflavored LG total $CP$ asymmetry can be defined as
\begin{eqnarray}\label{eq:epsilon}
 \epsilon_i = \frac{\sum_\alpha \left[ \Gamma(N_i\to e_\alpha\phi^\dag)-\Gamma(N_i\to \bar e_\alpha\phi) \right]}
	{\sum_\alpha \left[ \Gamma(N_i\to e_\alpha\phi^\dag)+\Gamma(N_i\to \bar e_\alpha\phi) \right]}.
\end{eqnarray}
Assuming for the couplings of scalar potential $\max( |\lambda_3|, |\lambda_4| ) \ll |\lambda_5|$ to suppress possible two-loop effects, 
the $CP$ asymmetry can be rewritten as~\cite{Fukugita_Yanagida,Strumia,StandardLGA,StandardLGB}
\begin{eqnarray}\label{eq:epsilon}
 \epsilon_i = \frac{1}{8\pi \left(YY^\dag\right)_{ii}}\sum_{j\neq1}{\rm Im}
\left[\left(YY^\dag \right)_{ij}^2\right] \, f \left(\frac{M_j^2}{M_i^2}\right),
\end{eqnarray}
where in EMM
\begin{eqnarray}
 f(x)= \sqrt{x}\left[ \frac{1}{1-x} + 1-(1+x)\ln\left(\frac{1+x}{x}\right)\right] 
= -\frac{3}{2\sqrt{x}} + \mathcal{O}(x^{-3/2}) \quad \text{for} \quad x\gg1,
\end{eqnarray}
and in EPWM
\begin{eqnarray}
 f(x)= \sqrt{x}\left[ 1-(1+x)\ln\left(\frac{1+x}{x}\right)\right] 
= -\frac{1}{2\sqrt{x}} + \mathcal{O}(x^{-3/2}) \quad \text{for} \quad x\gg1
\end{eqnarray}
since the only non-vanishing contribution comes from the vertex correction
~\cite{LG_BP1}.

The decay parameter can be written as
\begin{eqnarray}
 K \equiv \frac{\tilde\Gamma_{\text{D}}}{H|_{T=M_{\rho_3}}} = \frac{\tilde m}{m_*}, 
\end{eqnarray}
where $\tilde\Gamma_\text{D}$ is equal to the total decay rate $\Gamma_\text{D}$ in EMM and $\tilde\Gamma_\text{D} = \Gamma_\text{D}/3$ 
in EPWM, where it is normalized by the number of components of the triplet Majorana fermion.
The rescaled decay rate (effective neutrino mass) is defined as~\cite{Davidson_Nardi_Nir}
\begin{eqnarray}\label{eq:tilde_m}
 \tilde m \equiv 8\pi\frac{v_0^2}{M_{1}^2}\tilde\Gamma_{\text{D}} = \frac{v_0^2}{M_{1}} \left( YY^\dag \right)_{11},
\end{eqnarray}
and the rescaled Hubble expansion rate (equilibrium $N_1$ mass) is
\begin{eqnarray}
 m_* \equiv 8\pi\frac{v_0^2}{M_{1}^2}H|_{T=M_{1}} \simeq1.08\times10^{-3}~\text{eV}.
\end{eqnarray}
For NH (IH) the strong washout regime requires
\begin{eqnarray}
 K\geq K_{\text{sol}\,(\text{atm})}\equiv m_{2(1)}/m_*\simeq8.1\ (46)\gg1.
\end{eqnarray}

\subsubsection{Hierarchical $N_i$}

In the hierarchical limit $M_1/M_{i>1}\to0$, Eq.~\eqref{eq:epsilon} can be rewritten as 
\begin{eqnarray}\label{eq:epsilon1}
 \epsilon_1 = -\frac{A}{\left(YY^\dag\right)_{11}}  \sum_{j\neq1} \frac{M_1}{M_j} {\rm Im} \left[
\left(YY^\dag \right)_{1j}^2 \right]
  = -\frac{A M_1}{\left(YY^\dag\right)_{11}}  \Sigma
\end{eqnarray}
with $A=3/(16\pi)$ and $1/(16\pi)$ in the EMM and EPWM, respectively, and
\begin{eqnarray}\label{eq:Sigma}
  \Sigma  \equiv  \sum_{j\neq1} {\rm Im} \left[ 
\left(YY^\dag \right)_{1j}^2 M_j^{-1} \right] 
= \sum_{j=1,2,\dots} {\rm Im} \left[ 
\left(YY^\dag \right)_{1j}^2 M_j^{-1} \right].
\end{eqnarray}

Using Eqs.~\eqref{eq:Yprime} and \eqref{eq:Casas-Ibarra}, we have
\begin{eqnarray}\label{eq:YY}
   \left(YY^\dag\right)_{11} 
  = \frac{1}{v_0^2} \left( Q^T D_\mathcal{M}^{1/2} \Omega D_\nu \Omega^\dag D_\mathcal{M}^{1/2} Q \right)_{11} 
   = \frac{1}{v_0^2} \sum_\alpha m_\alpha \left| Q^T D_\mathcal{M}^{1/2} \Omega \right|_{1\alpha}^2.
\end{eqnarray}
Using Eq.~\eqref{eq:nu_masses:1}, we get 
\begin{eqnarray}\label{eq:epsilonSum}
  \Sigma = {\rm Im} \left[ \left(YY^\dag D_M^{-1} Y^*Y^T \right)_{11} \right]
  =  \frac{1}{v_0^2} {\rm Im} \left\{ \left[ Y(M_\nu - M_\nu^\text{loop})^\dag Y^T \right]_{11} \right\} 
  \equiv \Sigma_\nu + \Sigma_\nu^\text{loop},
\end{eqnarray}
where $\Sigma_\nu$ ($\Sigma_\nu^\text{loop}$) is the term with $M_\nu$ ($M_\nu^\text{loop}$).
Using Eqs.~\eqref{eq:Yprime}, \eqref{eq:Casas-Ibarra} and \eqref{eq:PMNS}, 
$\Sigma_\nu$ can be rewritten as
\begin{eqnarray}\label{eq:epsilonSum2}
   \Sigma_\nu = \frac{1}{v_0^4} {\rm Im} \left[ \left( Q^T D_\mathcal{M}^{1/2} \Omega D_\nu^2 \Omega^T D_\mathcal{M}^{1/2} Q \right)_{11} \right]
  = \frac{1}{v_0^4} \sum_\alpha m_\alpha^2 \, {\rm Im} \left[ \left( Q^TD_\mathcal{M}^{1/2}\Omega \right)_{1\alpha}^2 \right],
\end{eqnarray}
and, using Eqs.~\eqref{eq:nu_masses:1} and \eqref{eq:linear_relation}, $\Sigma_\nu^\text{loop}$ can be rewritten as
\begin{eqnarray}\label{eq:epsilonSum3}
  \Sigma_\nu^\text{loop} = - \frac{1}{\Lambda} {\rm Im} \left[ \left( Yh^* \right)_{1}^2 \right] 
  = - \frac{1}{\Lambda} {\rm Im} \left[ \left( YY^\dag a^\dag \right)_{1}^2 \right].
\end{eqnarray}

\vspace{1.5cm}

In the case of two $N_i$ we have
\begin{eqnarray}\label{eq:Sigma_2N}
  \Sigma  =  \frac{1}{M_2} {\rm Im} \left[ 
\left(YY^\dag \right)_{12}^2 \right]
\end{eqnarray}
and
\begin{eqnarray}\label{eq:epsilonSum3_2N}
  \Sigma_\nu^\text{loop} = - \frac{a_2^2}{\Lambda} \text{Im} \left[(YY^\dag)_{12}^2 \right] =  -a_2^2 \frac{M_2}{\Lambda} \Sigma.
\end{eqnarray}
From Eqs.~\eqref{eq:epsilonSum}, \eqref{eq:epsilonSum2} and \eqref{eq:epsilonSum3_2N}, we get
\begin{eqnarray}
   \Sigma = \frac{M_2^\prime}{M_2}\Sigma_\nu = \frac{1}{v_0^4} \frac{M_2^\prime}{M_2} \sum_\alpha m_\alpha^2 \, {\rm Im} \left[ \left( Q^TD_\mathcal{M}^{1/2}\Omega \right)_{1\alpha}^2 \right],
\end{eqnarray}
with
\begin{eqnarray}\label{eq:M2prime}
  M_2^\prime = \left( \frac{1}{M_2} + \frac{a_2^2}{\Lambda} \right)^{-1},
\end{eqnarray}
and, using Eqs.~\eqref{eq:epsilon1} and \eqref{eq:YY}, we have
\begin{eqnarray}\label{eq:epsilon_2N}
 \epsilon_1 \simeq   - A\mu \frac{M_1}{v_0^2} 
  \frac{ \sum_\alpha m_\alpha^2 \, {\rm Im} \left[ \left( Q^TD_\mathcal{M}^{1/2}\Omega \right)_{1\alpha}^2 \right] }
  {\sum_{\alpha} m_\alpha \left| Q^TD_\mathcal{M}^{1/2}\Omega \right|_{1\alpha}^2}.
\end{eqnarray}
We note that the magnification factor $\mu \equiv M_2^\prime/M_2$ is formally equivalent to the magnification of thin concave lens 
with the focal length $f=-\Lambda/a_2^2$ since Eq.~\eqref{eq:M2prime} can be rewritten as
\begin{eqnarray}
  \frac{1}{M_2} - \frac{1}{M_2^\prime} = \frac{1}{f}.
\end{eqnarray}

\subsection{Boltzmann equations}

Boltzmann equations in the unflavoured regime 
can be written as (for more details see~\cite{KZ,LG_BP1,LG_BP2,Davidson_Nardi_Nir,LG4pedestrians} and Refs. therein)
\begin{eqnarray}\label{eq:boltzmann_rho3}
 \frac{dN_{N_1}}{dz} &=& -(D+S)(N_{N_1}-N_{N_1}^{\rm eq}), \\ 
 \frac{dN_{B-L}}{dz} &=& -\epsilon_{1}D(N_{N_1}-N_{N_1}^{\rm eq})-WN_{B-L},
 \label{eq:boltzmann}
\end{eqnarray}
where $z=M_{1}/T$, and $N_X$ ($X=N_1,\,B-L$) is the number density of $X$ 
calculated in a co-moving volume containing one $N_1$ (all of its components) in ultrarelativistic thermal equilibrium: 
$N_{N_1}^{\text{eq}}(T \gg M_{1}) = 1$. Initially, $N_{B-L}^{\text{eq}}(T \gg M_{1}) = 0$.
$D=\Gamma_D/(Hz)$ is the decay factor, and $W$ is the washout term.
The scattering term is $S=S_\phi$ in EMM and $S=S_\phi+2S_g(N_{N_1}+N_{N_1}^{\rm eq})$ in EPWM (see footnote on p.~3), where $S_\phi$ is the 
contribution from Higgs-mediated scatterings, and the gauge scattering of the triplet Majorana fermion~\cite{Strumia} can be fitted by~\cite{LG_BP1}
\begin{equation}\label{eq:Sg}
  S_g\simeq 10^{-3}\frac{M_{\text{Pl}}}{M_{\rho_3}}\frac{\sqrt{1+\pi z^{-0.3}/2}}{(15/8+z)^2(1+\pi z/2)}e^{0.3z},
\end{equation}
where $M_{\text{Pl}}=1.221\times10^{19}$~GeV is the Planck mass.

After solving the system of Boltzmann equations~\eqref{eq:boltzmann_rho3}--\eqref{eq:boltzmann}, we obtain $N_{B-L}^{\text{f}} = N_{B-L} (z \to \infty)$ 
(in the calculations below the final value $z=10$ is used, where the fit in Eq.~\eqref{eq:Sg} is still applicable),
included in the final baryon asymmetry
\begin{equation}\label{eq:BEsolution}
  \eta_B \simeq 3\times 0.88\times 10^{-2} N_{B-L}^{\text{f}}.
\end{equation}
This result should be compared with the allowed values
\begin{equation}\label{eq:B-L_exp}
  5.1\times10^{-10}<\eta_B^{\text{BBN}}<6.5\times10^{-10}, 
\end{equation}
which come from the nucleosynthesis predictions and observed abundances of light elements~\cite{PDG2010}.

\subsection{Analysis}
\subsubsection{Non-resonant case}

For the particular spectrum $M_1\ll \min(|\Lambda|,M_2,|\Lambda+a_2^2M_2|) \equiv M_\text{min}$,  
using Eqs.~\eqref{eq:Mi_approx} and \eqref{eq:sinq_approx}, Eqs.~\eqref{eq:YY} and \eqref{eq:epsilonSum2} can be rewritten as
\begin{eqnarray}\label{eq:YYapprox}
   \left(YY^\dag\right)_{11} 
  = \frac{\tilde M_1}{v_0^2} \left[Q_{11}^2 \sum_\alpha m_\alpha|\Omega_{1\alpha}|^2 + \mathcal{O}\left(\sqrt{\frac{M_1}{M_\text{min}}}\right)\right],
\end{eqnarray}
\begin{eqnarray}\label{eq:Sigma_nu_approx}
   \Sigma_\nu =  \frac{\tilde M_1}{v_0^4} \left\{ Q_{11}^2 \sum_\alpha m_\alpha^2 \, {\rm Im} \left[(\Omega_{1\alpha})^2\right]
+ \mathcal{O}\left(\sqrt{\frac{M_1}{M_\text{min}}}\right)\right\}.
\end{eqnarray}
Hence the $CP$ asymmetry in Eq.~\eqref{eq:epsilon_2N} can be rewritten as
\begin{eqnarray}\label{eq:epsilon1approx}
 \epsilon_1 \simeq   - A\mu \frac{M_1}{v_0^2} \frac{ \sum_{\alpha} m_\alpha^2 {\rm Im} \left( \Omega_{1\alpha}^2 \right) }
  {\sum_{\alpha} m_\alpha \left| \Omega_{1\alpha} \right|^2} 
  =  - A \mu\frac{M_1}{v_0^2} \frac{(m_b^2-m_a^2) {\rm Im} \sin^2z}{m_a|\cos z|^2+m_b|\sin z|^2},
\end{eqnarray}
where $a=2\,(1)$ and $b=3\,(2)$ for NH (IH) neutrinos.
Eq.~\eqref{eq:epsilon1approx} results in the upper bound for the $CP$ asymmetry
\begin{eqnarray}\label{eq:epsilon1:max}
 |\epsilon_1| \lesssim   A \mu \frac{M_1}{v_0^2} (m_b-m_a),
\end{eqnarray}
which is equal to the DI bound~\cite{DI} (its non-supersymmetric version has the same factor $A$ as EMM), 
rescaled by $\mu$.
Using Eq.~\eqref{eq:YYapprox}, 
Eq.~\eqref{eq:tilde_m} can be rewritten as
\begin{eqnarray}\label{eq:tilde_m:new}
 \tilde m \simeq Q_{11}^2 \frac{\tilde M_1}{M_1} \sum_\alpha m_\alpha|\Omega_{1\alpha}|^2 \simeq 
m_a|\cos z|^2+m_b|\sin z|^2 \geq m_a,
\end{eqnarray}
which is the usual form.
In the considered non-resonant region the DI bound can not be significantly relaxed since $\mu\lesssim1$. However the new allowed parameter ranges for successful LG 
appear for large values of the decay parameter $K$ and for IH neutrino masses, as was shown for triplet LG in~\cite{KZ} (in the context of Adjoint $SU(5)$),
using precise formulas for $\tilde m$ and $\epsilon_1$ in the case of two $N_i$.

\begin{center}
 \begin{figure}[tb]
 \centering
\includegraphics[width=0.7\textwidth]{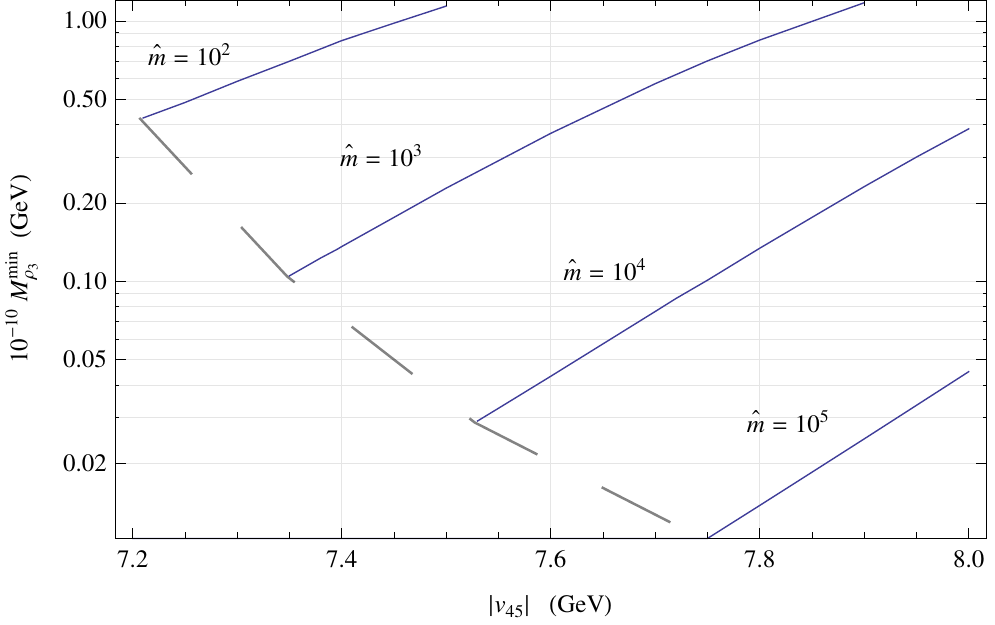}
   \caption{Minimal allowed by unflavoured LG values of $M_{\rho_3}$ vs. $|v_{45}|$ for $K=20$, $\lambda_5=-1/2$ and NH neutrinos in Adjoint $SU(5)$. 
  (Chosen values of $\hat m$ cover the region allowed by unification at 1-loop level, proton decay and collider searches~\cite{p-stability}.)
}
   \label{fg:Mmin}
 \end{figure}
\end{center}

\subsubsection{Resonant case}

For the case of approximate cancellation of the tree and loop contributions to the neutrino masses, namely $\Lambda\simeq -a_2^2M_2$ 
(see Eqs.~\eqref{eq:nu_masses:1} and \eqref{eq:linear_relation}), 
the factor $\mu$ is large and enhances the $CP$ asymmetry in Eqs.~\eqref{eq:epsilon_2N} and \eqref{eq:epsilon1approx}. 

In Adjoint $SU(5)$ model for $M_{S_8}\equiv M_{\eta}=1$~TeV and $\lambda_5=-1/2$ this resonance happens at $|v_{45}|\sim8$~GeV. 
By choosing values of $|v_{45}|$ near 8~GeV and using the method of calculations described in~\cite{KZ}, we get
minimal allowed by successful LG values of $\rho_3 \equiv N_1$ mass versus $|v_{45}|$, shown in Fig.~\ref{fg:Mmin} for $K=20$, NH neutrinos, SM Higgs mass $M_h=130$~GeV 
and several chosen values of $\hat m \equiv M_{\rho_8}/M_{\rho_3} \equiv M_N/M_{N_1}$
\footnote{We note that aside from the resonance area the maximal allowed value of $\hat m$ for unflavoured LG is $\sim10^4$~\cite{KZ}. 
However this value can be increased in the resonant case considered here.}. 
Below the dashed line in Fig.~\ref{fg:Mmin} the unflavoured LG is not allowed. 
Clearly, for stronger hierarchy of $M_i$ (higher values of $\hat m$) the lower bound on $M_{\rho_3}$ is weaker.
Fig.~\ref{fg:Mmin} shows that the allowed values of $M_{\rho_3}$ can be lowered by several orders of magnitude 
comparing to the scale of $10^{11}$~GeV, which is relevant for the case of vanishing loop contribution to the neutrino masses, see \cite{KZ,LG_BP1}.

In the case of singlet LG (as in EMM) the lower bound on the strongly hierarchical heavy neutrino masses (e.g., $M_2/M_1\gtrsim10^7$) can be decreased up to 
the TeV scale, which is testable at the LHC~\cite{heavy_neutrinosA,heavy_neutrinosB,ABZ11,ABZ12,ABZ21,ABZ22,ABZ3}. However it requires fine tuning of the parameters of the theory. 
We remark that this bound holds for the ordinary right-handed neutrinos 
in contrast to the reduced by the loop factor $16\pi^2$ DI bound on the masses of $Z_2$ odd Majorana fermions ($N$) derived in~\cite{Ma_LG}.

We note that the inert doublet model, which is embedded in the considered theory, 
may provide contribution to the dark matter in the universe, see~\cite{1106.1719} for recent study.


\section{Summary}\label{section:summary}

The SM extensions, which change the usual connection of the leptogenesis to the observable neutrino masses and relax the Davidson-Ibarra bound, are introduced. 
The lower bound on the hierarchical masses of heavy Majorana fermions can be significantly decreased, 
while the upper bound on the light neutrino masses may be increased in this theory, which may be tested in the near future experiments. 
The non-SM particles, involved in the loop contribution to the neutrino masses in Fig.~\ref{fig:2:loop:nu:masses}, such as scalar octet in Adjoint $SU(5)$ model
can be tested at the Large Hadron Collider and next colliders~\cite{Manohar-Wise,0809.2106,1010.5802}. 
Finally, the long standing gravitino problem can be solved in the supersymmetric version of this theory.

\section*{{Acknowledgments}}

The author thanks Riccardo Barbieri, Kristjan Kannike and Anatoly Borisov for useful discussions and comments, 
and the organizers of the BLV2011 Workshop Pavel Fileviez Perez and Yuri Kamyshkov for hospitality in Gatlinburg.
This work was supported in part by the EU ITN ``Unification in the LHC Era'',
contract PITN-GA-2009-237920 (UNILHC) and by MIUR under contract 2006022501.

\end{fmffile}

\end{document}